\documentclass[aps,prd,reprint,groupedaddress,showpacs]{revtex4-1}
\usepackage{amsmath,amssymb,graphicx,hyperref}

\newcommand{\eqn}[1]{\begin{eqnarray} #1 \end{eqnarray}}
\newcommand{\bk}[1]{\langle #1 \rangle}
\newcommand{\op}[1]{\hat{#1}}

\begin{document}

\title{Quantum fields on closed time-like curves}

\author{J.L. Pienaar}
 \email{j.pienaar@physics.uq.edu.au}
\author{C.R. Myers}
 \email{myers@physics.uq.edu.au}
\author{T.C. Ralph}
 \email{ralph@physics.uq.edu.au}
\affiliation{
 School of Mathematics and Physics, The University of Queensland, Brisbane 4072 QLD Australia
}

\date{\today}

\begin{abstract}
Recently, there has been much interest in the evolution of quantum particles on closed time-like curves (CTCs). However, such models typically assume point-like particles with only two degrees of freedom - a very questionable assumption given the relativistic setting of the problem. We show that it is possible to generalise the Deutsch model of CTCs to fields using the equivalent circuit formalism. We give examples for coherent, squeezed and single-photon states interacting with the CTC via a beamsplitter. The model is then generalised further to account for the smooth transition to normal quantum mechanics as the CTC becomes much smaller than the size of the modes interacting on it. In this limit, we find that the system behaves like a standard quantum mechanical feedback loop.
\end{abstract}


\maketitle

\section{Introduction}

The existence in general relativity of solutions that contain closed time-like curves has long posed a problem to physicists: can the laws of physics accommodate time travel without running into paradoxes\cite{MOR88,ORI05,CHI11}? The classic example is the science-fiction scenario in which a time-traveller kills his own grandfather, thus preventing his own existence and creating a paradox. The problem may be stated as 'what happens when we choose initial conditions such that the evolution on a CTC contradicts those initial conditions' ? The classic solution is an ad-hoc restriction on our freedom to choose initial conditions, known as the `Novikov consistency condition' or colloquially as the `banana peel mechanism' whereby we only allow initial conditions that contain a `banana peel' as a means of ensuring that any would-be grandfather killer will slip up and fail at his task\cite{NOV15}. Novikov's idea of placing constraints on the initial conditions was originally proposed in the context of classical general relativity, but it underlies many of the path integral approaches to quantum dynamics on CTCs due to Hartle, Politzer and others \cite{HAR94,POL94}. Although these attempts recognised that quantum mechanical effects can be significant and must be accounted for in CTC space-times, many of them relied upon a re-normalisation of the initial state in such a way as to exclude paradoxes - a procedure that is reminiscent of Novikov's proposal. As a result, it was found that severe problems remain in such theories, arising from the fact that the laws of physics in the past are altered due to the existence of a CTC in the future \cite{POL94}. The potential for new paradoxes is highlighted by the formulation of the traditional path integral approaches in terms of post-selection (P-CTCs)\cite{LLO11}, for which it has been argued that superluminal signalling outside the CTC epoch is a consequence\cite{RAL11}.

Deutsch\cite{DEU91} was the first to show that quantum mechanics might play a more fundamental role in resolving the paradoxes as demonstrated by his toy model of a chronology-respecting (CR) qubit (i.e. a point-like two-level system) interacting with another qubit trapped on a CTC. Instead of relying on a re-normalization of the CR-qubits initial state as in other approaches, Deutsch drew upon ideas from quantum information theory and proposed a consistency condition based on the density matrix of the CTC qubit. The resulting model places no constraints on the input state, always leads to a self-consistent solution (thereby solving the grandfather paradox) and does not affect the laws of physics prior to the CTC epoch, thereby avoiding many of the problems encountered in other models. The fundamental non-linearity of Deutsch's solution leads to increased power for certain quantum information tasks \cite{BAC04,BRU09}. Of particular relevance to this paper is the re-formulation of the Deutsch model in terms of `equivalent circuits'\cite{RAL10}.

  The equivalent circuit formulation of the model demonstrates that it is possible to keep track of the hidden degrees of freedom in the CTC and their correlations with the CR-qubit and thereby maintain coherence. In the equivalent circuit, the loss of coherence seen in the Deutsch model is interpreted as the tracing out of the inaccessible degrees of freedom by a detector in the asymptotic future. With this interpretation, the equivalent circuit reproduces the results of the Deutsch model. However, it also allows us to go beyond the Deutsch model in some important ways:
first, it explicitly includes the CTC degrees of freedom to provide a coherent unitary description, which we will use to replace Deutsch's simple qubits with more complex field states; second, it's structure suggests a further generalisation that allows us to treat fields whose wave-packets in space-time are of comparable size, or larger than, the CTC itself. In this way, we find a smooth transition between non-standard and standard quantum mechanics as a function of the temporal dislocation produced by the CTC. 

In section II we will review the Deutsch model and its equivalent circuit formulation. Section III is concerned with the extension of the model from two-dimensional states to those with an infinite dimensional Hilbert space. In section IV we discuss a nonlinear modification to quantum optics motivated by the equivalent circuit, which we use to perform the calculations of the previous section in situations where the wave-packets become comparable or larger than the CTC.  Our results are shown to be consistent with the Deutsch model in one limit and to recover standard quantum mechanics in another limit.

\section{Quantum circuit models of closed time-like curves}

\subsection{Some technical considerations}
 Unfortunately, space-times with CTCs do not admit foliation into a family of space-like hyper-surfaces, which is required for the notion of time evolution of a quantum field to be meaningful. Nevertheless, one might imagine that the CTCs are confined to a localised epoch in space and time, and hope to define some kind of scattering matrix between the asymptotic past and future. This is the approach used almost exclusively in the literature on the topic, and it will be used here. 

As emphasised by Hawking, all such attempts must result in a non-unitary scattering matrix if we insist on using quantum field theory in its accepted form\cite{HAW95}. This raises the problem of how to retain a probability interpretation, since non-unitary evolution seems to imply non-conservation of probability.

It is possible to evade this problem (as Hawking does) by pointing out that the loss of coherence comes from interacting with a part of the universe that is inaccessible insofar as it requires a theory of quantum gravity to describe its internal dynamics. Since we do not have such a theory, we are forced to trace it out, and we are justified in doing so as long as we only consider detectors in the far future that also cannot access the CTC. The problem of whether or not CTCs can be consistently described by physics is therefore postponed until we have a theory of quantum gravity. A related issue is Hawking's chronology protection conjecture\cite{HAW92}, which provides some indication that time travel will be impossible in a final theory; however it remains unconfirmed in the absence of such a theory.

It is nevertheless possible to ask: is there \textit{any} consistent way to modify the laws of physics to allow time-travel without paradoxes? If it can be shown that consistency of the theory necessarily implies that there is no observable time-travel, Hawking's conjecture would be confirmed. Alternately, if a consistent model of time-travel can be demonstrated through some modification of the laws of physics (whose predictions do not contradict the results of experiments), then the resulting theory may tell us something about quantum gravity.

In the next section we review an apparently consistent toy model of time travel, the Deutsch model, which we will subsequently generalise and extend to fields.

\subsection{The Deutsch model}
\begin{figure}[!htbp]
 \includegraphics[width=8.6cm]{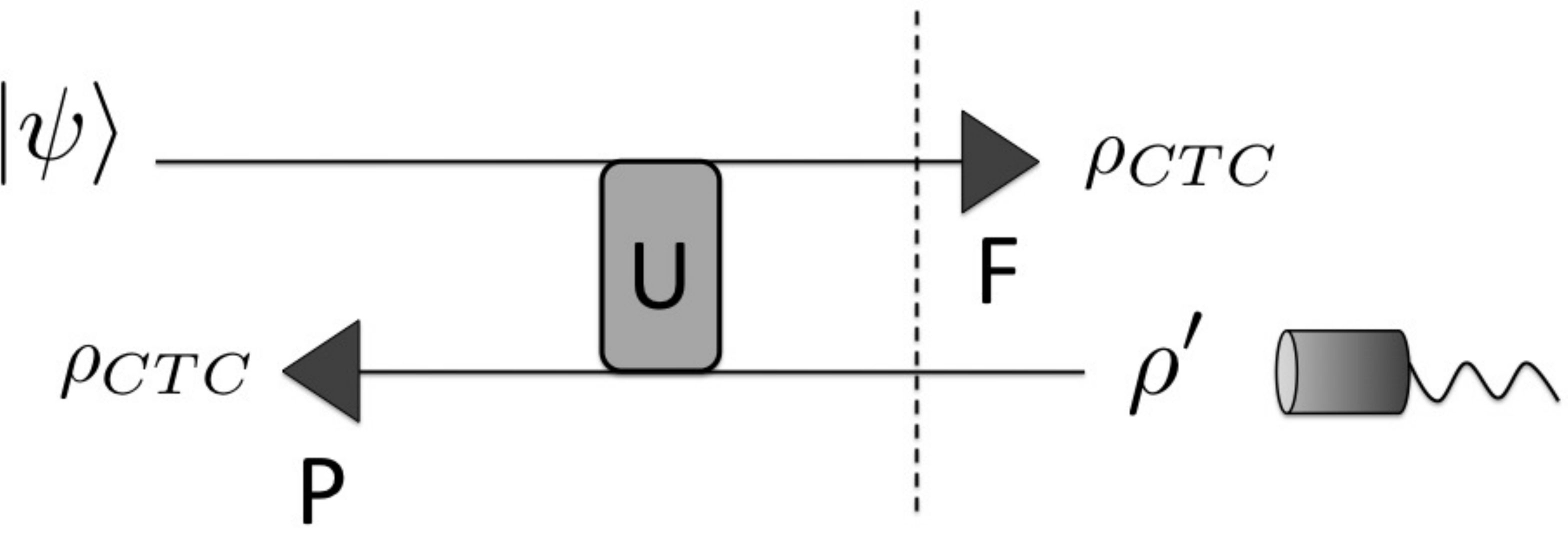}
\caption{The Deutsch circuit. An ingoing qubit $\rho$ interacts via a unitary U with it's time-traveling partner $\rho_{CTC}$. A consistency condition is applied to the reduced state $\rho_{CTC}$ entering the CTC in the future at `F' and emerging in the past at `P'. After solving this consistency condition, a trace is performed (signified by the dotted line) giving the reduced state $\rho'$ of the qubit at the detector.}
\label{deutsch}
\end{figure}

A qubit described by some density matrix $\rho = | \psi \rangle \langle \psi |$ (assumed to be pure) interacts via some two-qubit unitary $U$ with a second qubit in the unknown state $\rho_{CTC}$ which emerges from a CTC, as shown in Fig.\ref{deutsch}. The first qubit then enters the CTC, which sends it back in time, so that it undergoes the interaction again now playing the role of the second qubit. With this interpretation of the circuit, the interaction is seen to happen between the younger and older versions of the same qubit.

The state $\rho_{CTC}$ is constrained by the consistency condition:
\begin{equation}\label{cons}
 \rho_{CTC}=Tr_{\rho'} \left[ U \left( | \psi \rangle \langle \psi | \otimes \rho_{CTC} \right) U^\dagger \right]
\end{equation}
which can then be used to compute the evolution of $| \psi \rangle$:
\begin{equation}\label{deucirc}
\rho' = Tr_{CTC}\left[ U \left( | \psi \rangle \langle \psi | \otimes \rho_{CTC} \right) U^\dagger \right]
\end{equation}
where the final trace is performed over the CTC subsystem. The resulting input-output map is non-linear and non-unitary in general, but always has at least one solution for an arbitrary input and a given $U$. 

The Deutsch model, through the imposition of the constraint \ref{cons}, amounts to a non-linear modification to the laws of quantum mechanics. The nature of this non-linearity is different to that encountered in other models because of the nature of the constraint. As emphasised in \cite{WAL10}, by requiring matching of the density matrix rather than matching of the individual states in its pure-state decomposition (or equivalently matching of paths around the CTC), Deutsch implicitly treats the density matrix as an ontologically `real' object, not just as something representative of an observer's state of knowledge (an epistemic state). 

\subsection{The equivalent circuit}
The Deutsch model resolves the grandfather paradox without placing any constraints on the input state because the equation (\ref{cons}) always has at least one fixed point. However, for some specific interactions, there is more than one fixed point. To choose between them, Deutsch was forced to make an additional postulate that singles out the solution with the most entropy.
Fortunately, it is possible to re-formulate the model such that the maximum entropy principle emerges as a natural consequence of the dynamics. This re-formulation in terms of the `equivalent circuit' can be regarded as an extension of the Deutsch model that agrees with all of its predictions, but has some other nice properties as well. We review it briefly here; details can be found in \cite{RAL10}.

\begin{figure}[!htbp]
\includegraphics[width=8.6cm]{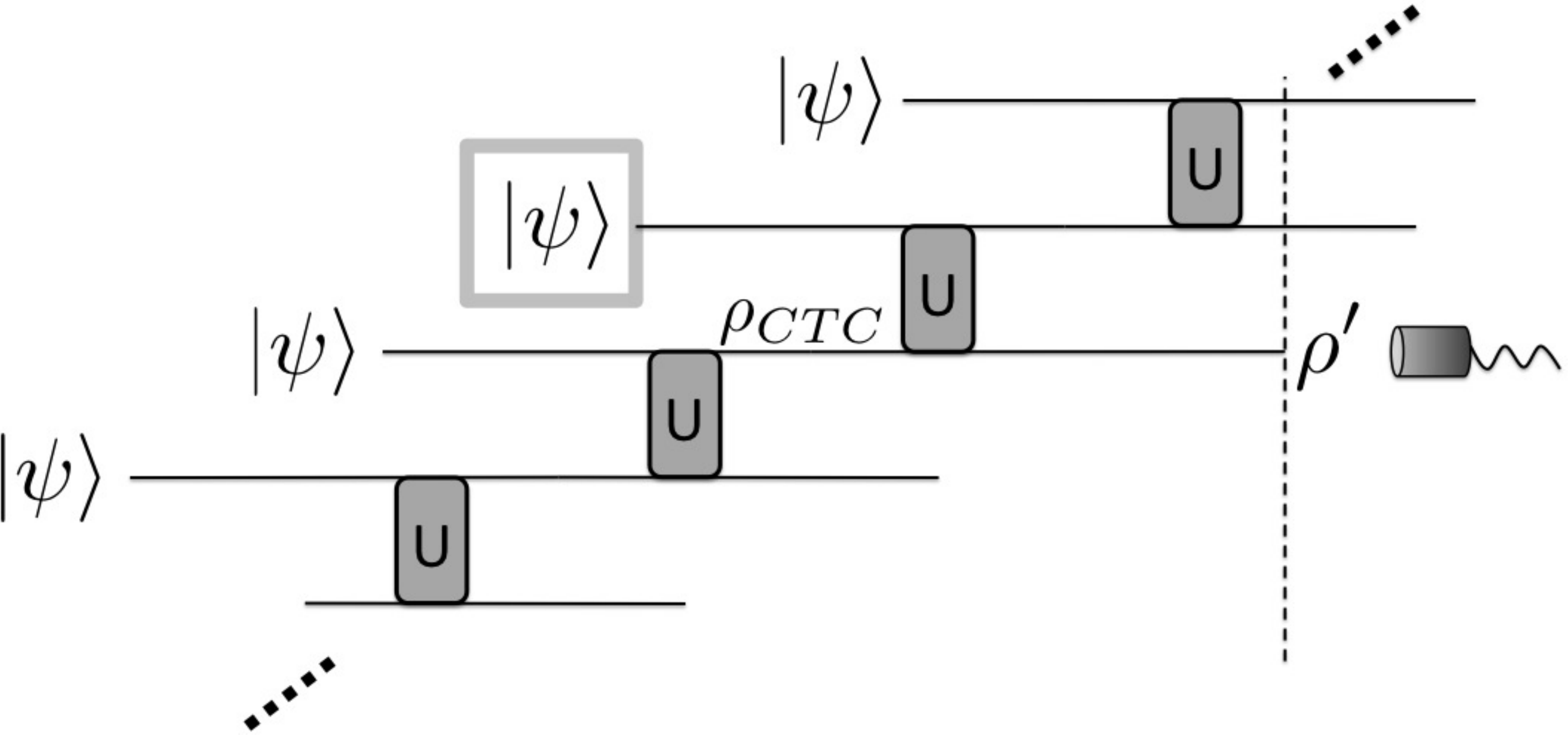}
\caption{The equivalent circuit corresponding to Fig.\ref{deutsch}. The consistency condition is now enforced by symmetry: the evolution of a large array of copies of the input state through a series of identical unitaries results in a fixed point for the output. A final trace is performed to obtain the reduced state of the outgoing qubit. The evolution up to the final measurement may be performed in the Heisenberg picture. While the predictions are the same as for the circuit of Fig.\ref{deutsch}, the equivalent circuit has a far richer Hilbert space.}
\label{equiv}
\end{figure}

To obtain the equivalent circuit from Deutsch's circuit, we consider the dynamics from the point of view of the time-traveling qubit. After passing through the interaction and entering the CTC on the top rail of the circuit, the qubit finds itself on the bottom rail of a new circuit, whose top rail is occupied by an identical copy of the qubit in its initial state. After the interaction, this copy then enters the CTC and goes on to interact with another copy, and so on. Formally, we replace Deutsch's circuit, Fig.\ref{deutsch}, with that of Fig.\ref{equiv}. 

The minimum size of the circuit is dictated by the number of iterations necessary to reach a fixed point. The existence of such a fixed point is guaranteed by the form of the map\cite{DEU91}. This number will be finite when there is just a single fixed point; when there are multiple fixed points, the introduction of a small amount of decoherence provides a means of obtaining a single fixed point deterministically after a finite number of iterations. Furthermore, this solution corresponds to the one selected by the maximum entropy postulate in Deutsch's model. Once the fixed point is found, a final trace is performed over the other outputs. The equivalent circuit therefore realises all of the predictions of the Deutsch model within the framework of a standard quantum circuit, which is advantageous because it allows the full quantum tool-box to be applied and provides a clear intuition about the behaviour of the CTC for a given choice of unitary. In particular, it is clear that the model is free from pathological behaviour such as global superluminal signalling, which appears in other models \cite{HAR94,RAL11}. The evolution is still non-linear, because of the presence of multiple copies of the initial state (in violation of the no-cloning theorem), and it is still non-unitary in general due to the presence of the final trace; therefore these intrinsic properties of the CTC are preserved by the framework. 

 The equivalent circuit also provides a clear intuition for extending the model to fields; we turn to this problem in the next section.

\section{Field operators}

To generalise the model, let the rails of the equivalent circuit represent the mode operators of a field of massless scalar bosons (e.g. the quantised electromagnetic field in 1-D) instead of point-like qubits. In quantum optics, we would associate each rail in this circuit with an operator of the form:
\begin{equation}
  \op{A}_{G} \equiv \int \mathrm{d}\textbf{k} \; G(k,x) \op{a}_{\textbf{k}}
\end{equation}
representing the annihilation of a photon in the wave-packet $G(k,x)$, whose Heisenberg evolution through the circuit can be used to define the dynamics. From here on without loss of generality we assume the wave-packet is a Gaussian superposition of plane waves in flat space: $G(k,x)=g(\textbf{k})e^{ikx}$, where $g(\textbf{k})$ is normalised and is equal to zero in the region $\textbf{k}<0$. Quantities of interest are then obtained by calculating the expectation values of the appropriate functions of this operator and its adjoint. 

As we are now explicitly incorporating the space-time co-ordinates into the problem, we should clarify what metric is being used for the CTC. For simplicity, following the example of Politzer\cite{POL94}, we consider fields evolving in flat space-time, where the CTC is implemented by making an identification between two space-like hyper-surfaces, one of which is in the causal past of the other. This is a highly contrived metric in which `traversal of the CTC' is treated as a pure temporal dislocation (formally defined in terms of translations of the equivalent circuit) with no additional dynamics or structure owing to the effects of extreme curvature that would normally be associated with a `realistic' metric. The results of section III will hold independently of such considerations given our assumptions of a large CTC and spatially localised modes, which imply that the space-time is locally flat along the particle paths. The validity of this argument will be re-examined in section IV, in which we seek to weaken the assumption of localised modes.

Returning to the equivalent circuit, each rail is now associated with a copy of the original field mode, differing only by some index `$m$' that delineates the extra degree of freedom of the CTC. Formally, we label the rails by $m$ running from $-\infty$ to $\infty$. We write:
\begin{equation}\label{sharpevent}
  \op{A}_{G,m} = \int \mathrm{d}\textbf{k} \;  g(\textbf{k})e^{ikx} \op{a}_{\textbf{k},m}
\end{equation}
for the mode associated to the $m_{th}$ rail. The modes satisfy the commutation relation:
\begin{equation}\label{comm1}
  [\op{a}_{\textbf{k},m},\op{a}^\dagger_{\textbf{k'},n}]=\delta(\textbf{k}-\textbf{k'})\delta_{m n}
\end{equation}
in accordance with our requirement that the different rails belong to different Hilbert spaces. This leads to the same-time wave-packet commutation relation:
\eqn{\label{precomm}
  \left.[\op{A}_{G,m},\op{A}^\dagger_{G,n}]\right|_{t=t'} \nonumber \\ 
  = \int \mathrm{d}\textbf{k} \; g(\textbf{k})g^*(\textbf{k}) \, e^{i \textbf{k}(\textbf{x}-\textbf{x}')} \delta_{m n} .
}
We will find it useful to consider the commutator between modes at the same point in space and time, obtained by setting $\textbf{x}=\textbf{x}'$ in (\ref{precomm}). We then obtain the `same-event' commutation relation:
\eqn{\label{comm2}  [\op{A}_{G,m},\op{A}^\dagger_{G,n}] = \delta_{m n}.}
This relation is the main point of departure from standard quantum mechanics, because it allows for the possibility of interactions between the rails of the equivalent circuit. Since these rails carry identical copies of the input state, interactions between them introduce the potential for nonlinear quantum behaviour characteristic of a CTC. Conversely, in the absence of interactions between the rails, the extra degree of freedom becomes degenerate and we obtain the commutator of normal, linear quantum mechanics independently on each rail; physically this is taken to correspond to decoupling from the CTC. Our goal is to give a detailed account of the different ways that this decoupling can happen, with reference to the particular example of a beamsplitter interaction.
For the rest of this section we maintain our assumption that the dimensions of the wave-packet $G(k,x)$ are much smaller than the CTC; we can then compute the output for any reasonable incident field mode, including those that contain superpositions of many particles. All we have to do is compute the evolution of the modes at the detector through the circuit in Fig.\ref{equiv} for a given $U$, and take the expectation values of the desired moments in the initial state $|\psi \rangle \equiv \bigotimes \limits^{\infty}_{m} |\psi_{m} \rangle$. This state represents an infinite number of copies of the original input state, which could be any state, such as a Fock state, a coherent state, or a squeezed state. We will give examples of all three in the remainder of this section, for the case where $U$ is a beamsplitter.

\subsection{The beamsplitter on a CTC}

\begin{figure}[!htbp]
\includegraphics[width=8.6cm]{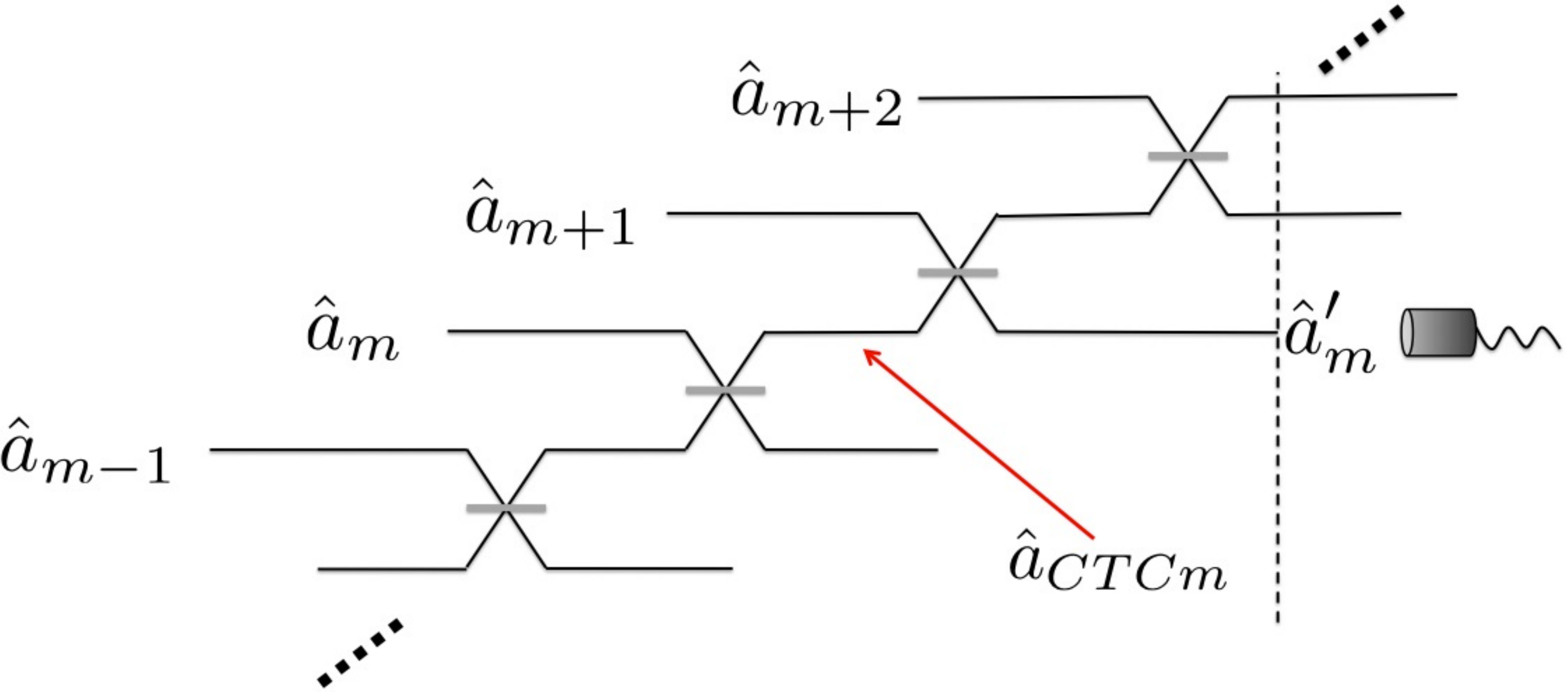}
\caption{(Colour online) The beamsplitter equivalent circuit. In the Heisenberg picture, we start with the mode being detected and evolve it through the unitary operator representing the infinite circuit. This produces an expression that is a function of the ingoing modes.}
\label{bs2}
\end{figure}

Consider the scenario in which the unitary in Fig.\ref{deutsch} is a beamsplitter; the equivalent circuit for this case is shown in Fig.\ref{bs2}. The Heisenberg evolution through this array is non-trivial because the circuit is formally infinite. Fortunately, we can make use of the graph's symmetry to compute the evolution by iteration.

Let the mode at the detector be $\op{a}'_m$. We wish to derive a general recipe for calculating it's evolution through the circuit. Following it back along the rail, we see that it satisfies:
\eqn{\label{aout} \op{a}'_m = U_{m+1} \, \op{a}_{CTC m} \, U^\dagger_{m+1} }
where $U_{m+1}$ is the two-mode unitary that acts on the $(m+1)$th and $m$th rails. The mode $\op{a}_{CTC m}$ is given by:
\eqn{\label{fdef} \op{a}_{CTC m} = U_{m} \op{a}_{m} U^\dagger_{m} \nonumber \\ \equiv f(\op{a}_{m},\op{a}_{CTC m-1})}
where the last step simply expresses the fact that output may be written as some function $f$ of the two input modes. Iterating this expression, we find:
\eqn{\label{actc} \op{a}_{CTC m} = f(\op{a}_{m},\op{a}_{CTC m-1}) \nonumber  \\ = f(\op{a}_{m},f(\op{a}_{m-1},\op{a}_{CTC m-2})) \nonumber  \\ = f(\op{a}_{m},f(\op{a}_{m-1},f(\op{a}_{m-2},...))).}
Given a specific unitary, we can determine $f$ and hence evaluate this expression for $\op{a}_{CTC\,m}$. Substituting the result into (\ref{aout}) will then give the output mode $\op{a}'_m$ in terms of the input modes.
We choose $U_{m}$ to be a linear beamsplitter, for which the Heisenberg evolutions for input modes $\op{a},\op{b}$ are:
\begin{equation}\label{beamsplit}
  \begin{aligned}
  \hat{a} &\rightarrow \sqrt{\eta}\, \hat{a}+e^{i\phi}\sqrt{1-\eta}\, \hat{b},\\
  \hat{b} &\rightarrow \sqrt{\eta}\, \hat{b}-e^{-i\phi}\sqrt{1-\eta}\, \hat{a},
 \end{aligned}
\end{equation}
where $\eta$ is the beamsplitter transmittance and $\phi$ is an arbitrary phase. Then: 
\eqn{&f(\op{a}_{m},\op{a}_{CTC m-1}) \nonumber \\ &= \sqrt{\eta} \, \op{a}_{m}+e^{i\phi} \sqrt{1-\eta}\, \op{a}_{CTC m-1}, }
and (\ref{actc}) evaluates to:
\eqn{ \op{a}_{CTC m} = \sqrt{\eta} \, \sum_{n=0}^{\infty} \, e^{i n \phi} (\sqrt{1-\eta})^{n} a_{m-n}}
Substituting this into (\ref{aout}), we obtain:
\eqn{ \op{a}'_m = \eta  \sum_{n=0}^{\infty} \,e^{i n \phi}  (\sqrt{1-\eta})^{n} a_{m-n} - e^{-i \phi}(\sqrt{1-\eta}) a_{m+1} \nonumber \\}
If we set $m=0$ and re-label the rails according to $\op{a}_x \rightarrow \op{a}_{-x}$, which we can do without loss of generality, the expression simplifies to:
\eqn{\label{aoutbs} \op{a}'_{0} = \eta  \sum_{n=0}^{\infty} \,e^{i n \phi}  (\sqrt{1-\eta})^{n} a_{n} - e^{-i \phi}(\sqrt{1-\eta}) a_{-1}. \nonumber \\}

\subsection{The coherent state}
The Heisenberg evolution for a coherent state of amplitude `$\alpha$' is:
\eqn{\label{coh} \op{D}(\alpha) \op{a} \op{D}^\dagger(\alpha) = \op{a} + \alpha}
where $\op{D}(\alpha)$ is the unitary displacement operator. The initial state is the vacuum: 
\eqn{ \label{vac} |0 \rangle = ...|0 \rangle_{1} \otimes |0 \rangle_{2} \otimes ... \otimes |0 \rangle_{\infty}.}
To calculate the various moments of $\op{a}'_0$ at the detector, we perform the Heisenberg evolution through the equivalent circuit given by (\ref{aoutbs}) followed by the preparation unitary (\ref{coh}) and then take expectation values in the vacuum state using the commutation relation (\ref{comm2}). Since the evolution consists entirely of Gaussian operations and inputs, we expect the output to also be Gaussian and therefore we can   characterise the state by its first- and second-order moments\cite{BAC}.
For the coherent state, we find:
\eqn{ \label{coh2}
\bk{\op{a}'} = \frac{1-e^{-i\phi}\sqrt{1-\eta}}{1-e^{i\phi}\sqrt{1-\eta}}\alpha \equiv \gamma, \nonumber \\
\bk{\op{a}'^\dagger \op{a}'} = |\gamma|^2 = |\alpha|^2, \nonumber \\
  \bk{\op{P}} \equiv \bk{\frac{i}{\sqrt{2}}(\op{a}'^\dagger-\op{a}')} =1/\sqrt{2}(\gamma^*-\gamma), \nonumber \\
  \bk{\op{Q}} \equiv \bk{\frac{1}{\sqrt{2}}(\op{a}'^\dagger+\op{a}')} = 1/\sqrt{2}(\gamma^*+\gamma), \nonumber \\
  Var\bk{\op{P}} = Var\bk{\op{Q}} = \frac{1}{2}.
  }

Noting that $\gamma \equiv e^{i \Phi(\eta,\phi)} \alpha$, together with the results (\ref{coh2}), shows the output of the equivalent circuit is again a coherent state, with the added phase $\Phi(\eta,\phi)$. This is just standard quantum mechanics; to a coherent state, a CTC looks no more strange than a simple unitary phase shift. To explain the intuition behind this result, we recall that the beamsplitter interaction between two coherent states belongs to a class of problems in which the unitary does not produce entanglement between the outputs. This means that the function in (\ref{fdef}) has the form: 
\eqn{  f(\op{a}_{m},\op{a}_{CTC m-1}) = \op{a}_{CTC m-1} }
which implies that the interaction can be written as:
\eqn{ \op{U}_{m} = \op{u}_{m} \op{U}_{mSWAP} } 
where $\op{u}_{m}$ acts only on the $m_{th}$ rail and $\op{U}_{mSWAP}$ swaps the $m_{th}$ and $(m-1)_{th}$ rails. Substituting this into (\ref{aout}) gives:
\eqn{ \op{a}'_m = \op{u}_{m+1} \, \op{a}_{m+1} \, \op{u}^\dagger_{m+1} }
which clearly represents ordinary quantum mechanics and is independent of the CTC. Therefore in all situations where the beamsplitter does not generate entanglement between the input mode and the CTC mode, the dynamics is decoupled from the CTC and normal quantum mechanics is restored. We are immediately led to wonder what might occur should the beamsplitter generate entanglement. One such example is the case of a squeezed vacuum input, to which we now turn.

\subsection{Squeezed vacuum}
The unitary squeezing operator $\op{S}(\xi)$ is defined as\cite{WAL}:
\eqn{\label{squin} \op{S}(\xi) \equiv e^{\frac{1}{2}(\xi^* \op{a}^2-\xi (\op{a}^\dagger)^2)}, }
where the polar decomposition $\xi \equiv r e^{-i 2 \theta}$ conventionally defines the squeezing parameter $r > 0$ and squeezing angle $\theta$. 
The squeezing of a vacuum mode $\op{a}$ is given by the Heisenberg evolution:
\begin{equation}
 \op{S}(\xi) \, \op{a} \, \op{S}^\dagger(\xi) = cosh(r)\op{a}-e^{-i2\theta}sinh(r)\op{a}^\dagger.
\end{equation}
For a squeezed vacuum input, the initial state is again the vacuum state and (\ref{squin}) is the preparation unitary. Proceeding as before, we evolve the modes at the detector through the circuit back to the vacuum and we obtain the following results from the first and second order moments:
\eqn{ \label{squeezed} 
  \bk{\op{a}'} = 0, \nonumber \\
  \bk{\op{a}'^\dagger \op{a}'} = sinh^2(r), \nonumber \\
  \bk{\op{P}} = \bk{\op{Q}} = 0, \nonumber \\
  Var\bk{P} = \frac{1}{2}\left( cosh(2r)-L(\phi,\theta,\eta)\,sinh(2r) \right) \nonumber \\
  Var\bk{Q} = \frac{1}{2}\left( cosh(2r)+L(\phi,\theta,\eta)\,sinh(2r) \right),
}
where
\begin{equation}\nonumber
 \begin{aligned}
  &L(\phi,\theta,\eta) \equiv \frac{(1-2\eta)cos(2\theta)+(\eta-1)^2M(\theta,\phi)}{2+(\eta-2)\eta+2(\eta-1)cos(2\phi)},
 \end{aligned}
\end{equation}
and
\eqn{M(\theta,\phi) \equiv cos(2\theta+4\phi)-2cos(2\theta+2\phi).}

\begin{figure}[!htbp]
\includegraphics[width=8.6cm]{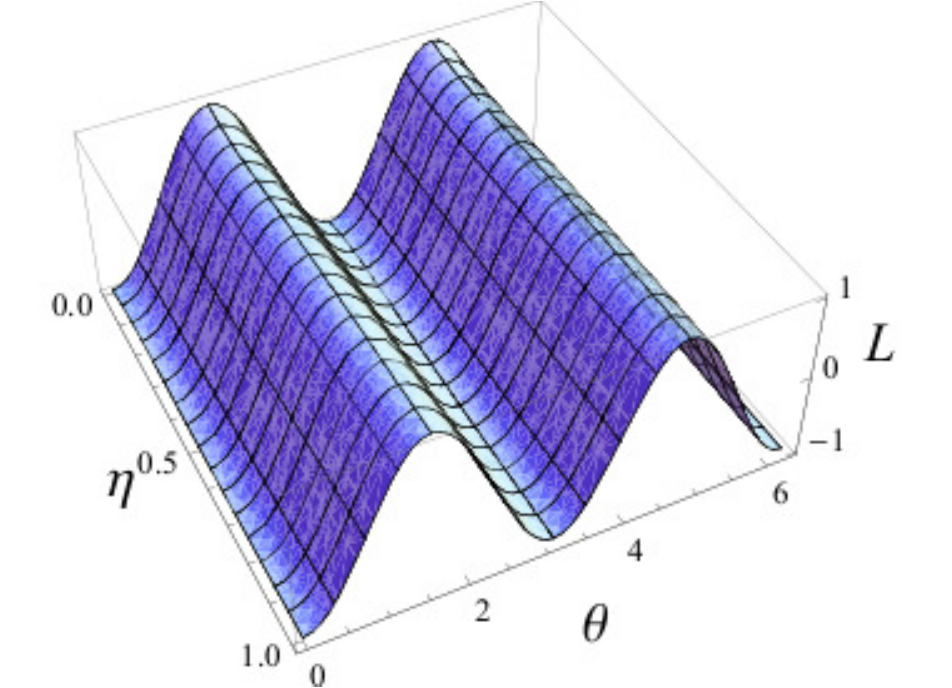}
\caption{(Colour online). Plot of $L(0,\theta,\eta)$. We see that $|L(0,\eta)|_{max} = 1$ regardless of $\eta$, so there is no added noise. Furthermore there is no rotation of the squeezing angle since $\theta$ is constant with $\eta$. Intuitively this occurs because $\phi=0$ means that the squeezed states always combine in phase on the beamsplitter, leading to no entanglement generation.}
\label{phi0}
\end{figure}

\begin{figure}[!htbp]
\includegraphics[width=8.6cm]{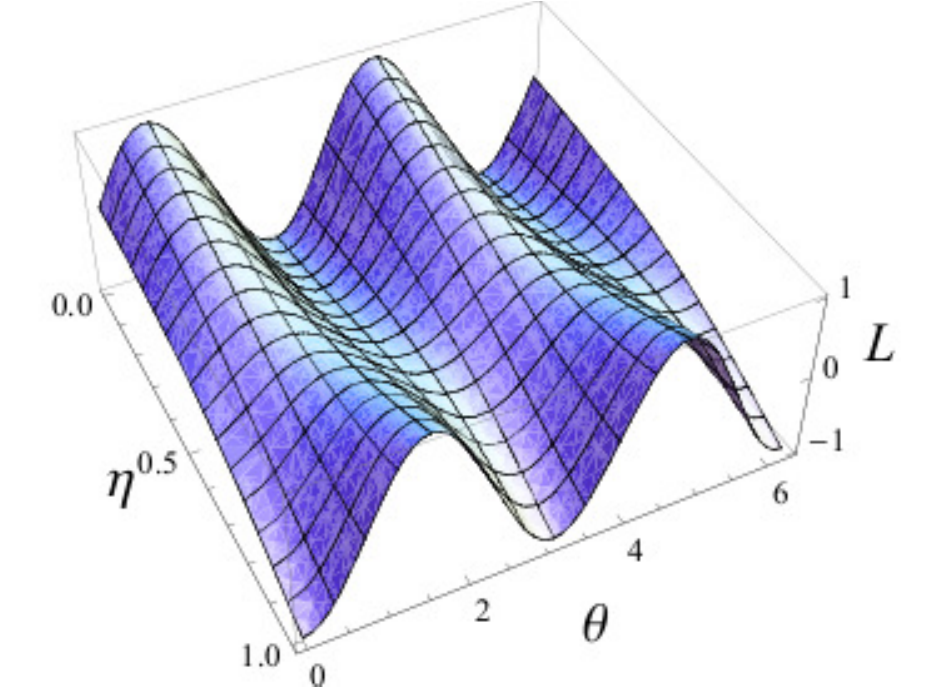}
\caption{(Colour online). Plot of $L(\frac{\pi}{4},\theta,\eta)$. Some noise is present for $0<\eta<1$, and $\theta$ rotates as a function of $\eta$ up to $\theta+\frac{\pi}{4}$ at $\eta=0$.}
\label{phi1}
\end{figure}

\begin{figure}[!htbp]
\includegraphics[width=8.6cm]{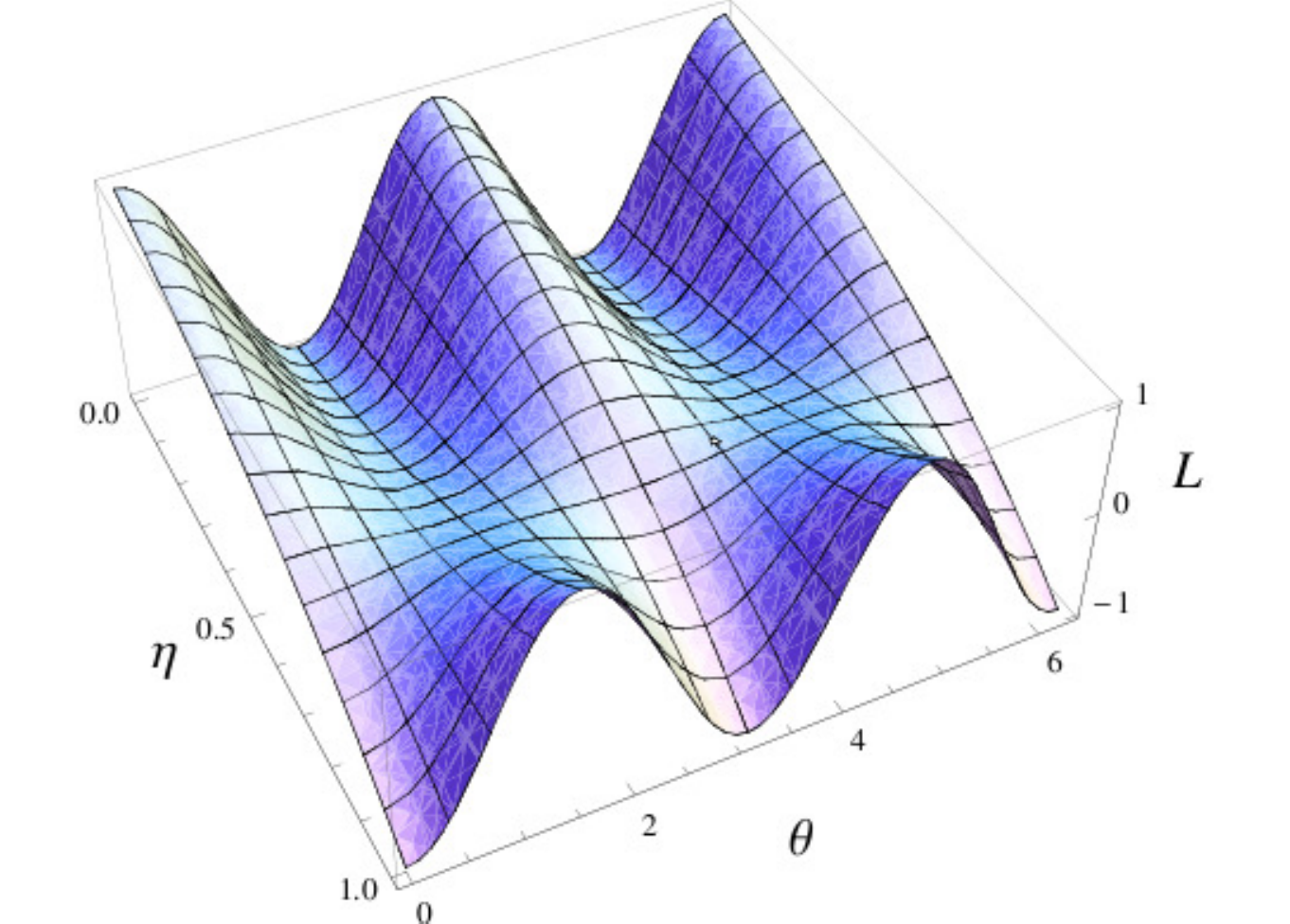}
\caption{(Colour online). Plot of $L(\frac{\pi}{2},\theta,\eta)$. When $\eta=\frac{2}{3}$ we obtain the maximum noise: $|L(\frac{\pi}{2},\frac{2}{3})|_{max} = 0$. This is again accompanied by a rotation of $\theta$ by up to $\phi=\frac{\pi}{2}$. }
\label{phi2}
\end{figure}

The function $L(\phi,\theta,\eta)$ takes real values between -1 and 1. If $L(\phi,\theta,\eta)= cos(2\theta)$ we recover the result characteristic of a squeezed state with squeezing parameter $r$ and squeezing angle $\theta$, i.e. the input state; such a result would indicate that the state is completely unaffected. Unsurprisingly, we find that this occurs trivially whenever $\eta=1$, signifying total transmission. In the limit of total reflection, $\eta=0$, we find that $L= cos(2\theta+2\phi)$, signifying a rotation of the squeezing angle by $\phi$, so again the CTC has no effect up to a phase. 
However, in between these two limits, the CTC will not just rotate the state but will introduce noise as a function of $\phi$ and $\eta$. To characterize the noise, we consider the maximum of the absolute value of $L$, where the maximisation is taken over $\theta$ with $\phi$ and $\eta$ remaining free parameters; we will denote this quantity $|L(\phi,\theta_{max},\eta)| \equiv |L(\phi,\eta)|_{max}$ where $\theta_{max}(\phi,\eta)$ is the (not necessarily unique) value of $\theta$ that maximises $|L(\phi,\theta,\eta)|$ for a given $\phi$ and $\eta$. It has the property that whenever there is no added noise, there will always be some $\theta_{max}$ such that $|L(\phi,\eta)|_{max}=1$. In the graphs Figs.\ref{phi0}-\ref{phi2} (colour online), the presence of noise can be determined by noting whether the oscillations along the $\theta$ axis are subject to damping; if they are not, then there is no noise.

In the first figure, $\phi=0$,  and there is no rotation of the squeezing angle $\theta$, nor is there any added noise. In the last figure, the most extreme case, $\phi=\pi/2$ and there is a rotation of the squeezing angle by an amount that ranges from 0 to $\pi/2$ as $\eta$ decreases from 1 to 0. Noise is present for $0<\eta<1$, with maximum noise occurring at $\eta=\frac{2}{3}$, in which case the state is symmetric and completely thermal (as indicated in Fig.\ref{phi2} by the fact that $|L(\phi,\eta)|_{max}=0$ there). For intermediate values of $\phi$, the centre figure shows that the rotation of $\theta$ ranges from 0 to $\phi$, with the amount of rotation increasing as $\eta$ decreases, and the overall noise decreases as $\phi$ decreases, becoming zero when $\phi$ also vanishes.

In summary, the effect of the CTC is to add noise to the squeezed vacuum and to change the angle of squeezing. Both effects occur whenever $\phi \neq n 2 \pi, n=0,1,2...$ and $0<\eta<1$, otherwise neither effect is seen. 

\subsection{Single photon state}

When the input is a single photon, the usual approach would be to place a one-photon Fock state $|1\rangle$ on each rail of the equivalent circuit and take expectation values in the initial state $|\psi \rangle \equiv \bigotimes \limits^{\infty}_{m} |1\rangle_{m} $. This is possible because all the results presented in this section can in principle be derived in the Schr\"odinger picture. However, we will find it useful for our considerations in section IV to take expectation values in the vacuum state (\ref{vac}) as we have done for the coherent and squeezed states; for this reason we adhere to treating the single photon state in the Heisenberg picture right down to the vacuum. The Heisenberg evolution of a vacuum mode $\op{a}$ to a mode containing exactly one photon is described by a unitary single-photon source, the details of which may be found in \cite{PIE11}. We need not reproduce the full expression here, but only take note of the following useful properties derived from it for a one-photon mode in the vacuum:
\begin{equation} \label{identities}
 \begin{aligned}
  & \langle 0|\hat{a} |0\rangle = 0, \\
  & \langle 0|\hat{a}^\dag \hat{a} |0\rangle = 1, \\
  & \langle 0|\hat{a}^\dag \hat{a}^\dag \hat{a} \hat{a}|0\rangle=0, \\
  & [\hat{a},\hat{a}^\dag] = 1 \, .
 \end{aligned}
\end{equation}
Proceeding as before, we use (\ref{aoutbs}) to write the output in terms of the input modes and then (\ref{identities}) to take the expectation values in the vacuum state. We then find that the output of the CTC has the following moments:
\eqn{ \label{sing1} \bk{\op{a}'^\dagger \op{a}'} &=& 1,}
and
\eqn{ g^{(2)} &=& \bk{\op{a}'^\dagger \op{a}'^\dagger \op{a}' \op{a}'} \nonumber \\
&=&4\eta(1-\eta) \\
&&+ \,\sum_{n,m,p,q=0}^{\infty} \eta^4(\sqrt{1-\eta})^{n+m+p+q} \langle a^\dagger_{n}a^\dagger_{m}a_{p}a_{q} \rangle. \nonumber 
 }
To evaluate this quantity, we note that (\ref{identities}) implies:
\eqn{
\bk{ a^\dagger_{n}a^\dagger_{m}a_{p}a_{q} } = \begin{cases} 1 & \textrm{if } \, (n=q)\neq(m=p), \\
 & \textrm{or } \, (n=p)\neq(m=q), \\
 0 & \textrm{otherwise}
\end{cases}
}
leading to the equivalent expression:
\eqn{\label{nmpq} &\sum \limits_{n,m,p,q=0}^{\infty}& (\sqrt{1-\eta})^{n+m+p+q} \bk{ a^\dagger_{n}a^\dagger_{m}a_{p}a_{q} } = \nonumber \\ &\sum \limits_{n,m,p,q=0}^{\infty}&  (\sqrt{1-\eta})^{n+m+p+q} \nonumber \\
&& \times \left( \delta_{nq} \delta_{mp} + \delta_{np} \delta_{mq} - 2\, \delta_{nm}\delta_{np}\delta_{nq} \right). \nonumber \\ }
This gives us the result:
\eqn{\label{g21} g^{(2)}=\frac{8\eta(\eta-1)}{\eta-2}.}

The result (\ref{sing1}) indicates that the average photon number is conserved as expected. For a single-photon state however, we would also expect $g^{(2)}$ to vanish; while this occurs for perfect reflection or transmission, $\eta=0,1$, it is not the case for values of $\eta$ in between these limits. In fact, we find from (\ref{g21}) that $g^{(2)}$ has a maximum of $24-16\sqrt{2} \approx 1.37...$ at $\eta=\frac{\sqrt{7}-1}{3}\approx 0.55$. This represents the value of the reflectivity for which the added noise is a maximum. We note that $g^{(2)}$ has sub-Gaussian statistics, since the kurtosis of the distribution $g^{(2)}(\eta)$ is found to be $\frac{2}{3}g^{(2)}+\frac{5}{3}$, which is less than the Gaussian result of $3$, for all $\eta$. These observations indicate that the single photon state becomes mixed by the CTC, but never completely thermalised, unlike the squeezed state.

\section{Generalisation of the equivalent circuit}

In the preceding calculations the particular shape of the wave-packet did not play any role, as it was assumed to be localised in space to a region much smaller than the scale of the CTC. We have seen that in spite of this restriction, different choices of input states led to varying amounts of coupling with the CTC, and some, like the coherent state, did not couple to the CTC at all, collecting only a phase shift. We would now like to relax our initial assumption and ask whether our model can be extended to situations in which the modes are longer than the CTC, such that the nose of the wave-packet could be sent back in time while the tail was still far away. We would expect that very long modes might decouple from the CTC due to an effective limitation on the interaction allowed between the mode and its time-travelling parts. In particular, for long modes we expect the noise observed  in section III, in the cases where the unitary was entangling, to disappear. In this section, we will modify the model of the previous section to take the effects of extended wave-packets into account.

Before we continue, we should address the issue raised in section III regarding the role of curvature in our model. Thus far we have justified ignoring the effects of curvature due to the CTC in two ways: first, we avoid the problem of defining modes in a non-globally hyperbolic space by treating the formalism strictly as an input-output map between asymptotically flat space-times; second, we disregard the interplay between the spatial properties of the modes and the CTC itself by assuming that the modes are spatially localised wave-packets much smaller than the CTC. It is the latter assumption that we now wish to relax, and it could be argued that this calls into question the validity of our model.
For this reason, the considerations of this section should be treated as a tentative starting point for the construction of a more complete model that would include curvature, and the results derived here as indications of the qualitative behaviour we might expect from such a model. In generalising the equivalent circuit in this section, we will find it necessary to postulate a connection between the spatial and CTC degrees of freedom that could point the way for future research into the role of curvature in this model.

Returning to the equivalent circuit, we have not yet suggested any physical interpretation for the extra degree of freedom. To extend the model, such an interpretation will prove useful; let us then consider Fig.\ref{deutsch} in the special case where the interaction is the identity. We now find that the incoming mode travels back in time and then escapes to the detector without any interaction with its younger self. In the equivalent circuit, this scenario involves nothing more than the detection of an `older copy' of the original mode instead of the mode itself. Thus, an incrementation of the parameter `$m$' can be interpreted as an `ageing' of the mode. This ageing ought to be quantified by an invariant parameter depending only on the length of the particle's path through the CTC. We therefore select some affine parameter $\tau$ that parameterises the particle's world line to keep track of the particle's `age' as we have interpreted it; then the incrementation of $m$ by some number $n$ corresponds to a shift in the mode of $n \Delta \tau$ along the world-line.
This reasoning suggests we replace our model with a more general model in which the discrete index `$m$' is replaced by a continuous parameter $\Omega$, whose Fourier complement is $\tau$. Then we are led to replace Eq.\ref{sharpevent} with the new wave-packet operator:
\begin{equation}\label{eventop0}
  \op{A}_{G,J} \equiv \int \mathrm{d}\textbf{k} \,  g(\textbf{k}) e^{ikx} \int \mathrm{d} \Omega \,  J(\Omega,\tau) \op{a}_{\textbf{k},\Omega}.
\end{equation}
We will choose the distribution in $\Omega$ to be a Gaussian: $J(\Omega,\tau) = j(\Omega)e^{i\Omega \tau}$. This choice will be justified from physical arguments when we consider the beamsplitter example; for the moment we take it as just a mathematical convenience. These generalised modes are subject to the commutation relation:
\begin{equation}\label{eocomm}
  [\op{a}_{\textbf{k},\Omega},\hat{a}^\dagger_{\textbf{k'},\Omega'}]=\delta(\textbf{k}-\textbf{k'})\delta(\Omega-\Omega').
\end{equation}
We note that a formally equivalent expression to (\ref{eventop0}) was derived by the authors of \cite{RAL09} through similar considerations; the relevance of that work is discussed at the end of this section. Since traversal of the CTC `n' times (or equivalently, a translation of `n' rails in the equivalent circuit) produces the transformation $J(\Omega, \tau) \rightarrow J( \Omega, \tau+n \Delta \tau)$ in our model, we write down the resulting transformed mode as:
\begin{equation}\label{shiftedop}
  \op{A}_{(n)} \equiv \int \mathrm{d}\textbf{k} \,  g(\textbf{k}) e^{ikx} \int \mathrm{d} \Omega \,  j(\Omega) e^{i \Omega (\tau+n\Delta \tau)} \op{a}_{\textbf{k},\Omega}
\end{equation}
It follows from (\ref{eocomm}) that the `same-event' commutator between a mode that has traversed the CTC $n$ times and a mode that has traversed the CTC $m$ times is:
\begin{equation}\label{commJ}
  \begin{aligned} 
    \left[ \op{A}_{(n)}, \op{A}^\dagger_{(m)} \right] & =\int \mathrm{d}\Omega\; |j(\Omega)|^2 \, e^{i \Omega(m-n)\Delta \tau} \\
     & =e^{-\frac{(m-n)^2\Delta \tau^2 \sigma^2_j}{4}} \equiv C_{n,m}.
 \end{aligned}
\end{equation}
We see that this decays exponentially as the difference $n-m$ increases. The rate of decrease of $C_{n,m}$ is controlled by the ratio $\kappa \equiv \frac{\Delta \tau}{\tilde{\sigma}_j}$ where $\tilde{\sigma}_j$ is the variance of $\tilde{j}(\tau)$, the Fourier transform of $j(\Omega)$. We see that when the shift $(m-n)\Delta \tau$ is much larger than the variance along $\tau$, the commutator $C_{n,m}$ will vanish when $n\neq m$ and we recover (\ref{comm2}), giving us the equivalent circuit. If, however, the commutator $C_{n,m}$ is nonzero for $n \neq m$, the interaction will be partly decoupled from the CTC. In that case we expect behaviour that asymptotes smoothly between the equivalent circuit of section III and standard quantum optics for which $C_{n,m} \rightarrow 1$. The new mode (\ref{eventop0}) therefore provides the machinery we need to describe what happens when the modes become larger than the scale of the CTC. To proceed further, we need to establish a connection between the function $\tilde{j}(\tau)$ that determines the coupling to the CTC degree of freedom and the wave-packet $\tilde{g}(\textbf{x},t)$ that defines the spatial properties of the mode. The nature of this connection becomes clear when we consider the beamsplitter example.

\subsection{The beamsplitter revisited}
It follows from (\ref{shiftedop}) and (\ref{aoutbs}) that the output of the CTC is now given by:
\eqn{\label{aoutbs2} \op{A}'_{(0)} = \eta  \sum_{n=0}^{\infty} \,e^{i n \phi}  (\sqrt{1-\eta})^{n} \op{A}_{(n)} - e^{-i \phi}(\sqrt{1-\eta}) \op{A}_{(-1)}. \nonumber \\ }
and we replace the `sharp' commutator (\ref{comm2}) of the equivalent circuit with the generalised commutator (\ref{commJ}). We now examine the dependence of our earlier results on the parameter $\kappa$ to see what happens when the modes are made longer or shorter (in the $\tau$ direction) compared to the CTC. Calculations for the general case are nontrivial, but for the limit of very long modes, $\kappa \rightarrow 0$, we find that $\op{A}_{(m)}=\op{A}_{(n)}\equiv \op{A}$ and (\ref{aoutbs2}) becomes the trivial evolution:
\eqn{\label{aoutstd} \op{A}' &=& \left( \eta  \sum_{n=0}^{\infty} \,e^{i n \phi}  (\sqrt{1-\eta})^{n} - e^{-i \phi}(\sqrt{1-\eta}) \right) \op{A} \nonumber \\
&=& e^{i \Phi(\eta,\phi)} \op{A}
}
This corresponds to the limit in which the mode does not `see' the CTC due to the variance of $\tilde{j}(\tau)$ being very large. We observe that this is the same result obtained in section III in the cases where the CTC became decoupled. However, the derivation leading to (\ref{aoutstd}) suggests a physical interpretation for the phase shift observed in such cases: since the evolution now appears to involve just a single mode cycling through the unitary (instead of multiple copies of the mode), we note that (\ref{aoutstd}) has the same form as a zero-delay feedback loop, because the transformed mode at the second input is defined by the same operator as the first input mode \cite{COM}. To take this reasoning further, note that a large feedback loop reduces to this same limit when the cross-section along $\tau$ of the spatial wave-packet $\tilde{g}(\textbf{x},t)$ is very long compared to the size of the delay. Hence the limit of very long $\tilde{j}(\tau)$ in the CTC model coincides with the limit of very long $\tilde{g}(\tau)$ on an ordinary feedback loop with length parameterised by $\tau(\textbf{x},t)$. The simplest way to account for this coincidence is to identify $\tilde{j}(\tau)$ with $\tilde{g}(\tau)$ and therefore $j(\Omega) \equiv g(\Omega)$ for the Fourier transformed modes. This justifies choosing $J(\Omega,\tau)$ to be Gaussian whenever the spatial modes are Gaussian. We take this as a postulate to connect the CTC coupling to the spatial properties of the wave-packet, leading us to replace (\ref{eventop0}) with:
\begin{equation}\label{eventop}
  \op{A}_{G,J} \equiv \int \mathrm{d}\textbf{k} \,  g(\textbf{k}) e^{ikx} \int \mathrm{d} \Omega \,  g(\Omega)e^{i\Omega \tau} \op{a}_{\textbf{k},\Omega}.
\end{equation}
The implications of this postulate will be discussed at the end of this section; for the moment we merely use it as a tool for fixing $j(\Omega)$ in our calculations.

\subsection{Wave-packet decomposition}
For the general case, we expect the output of the CTC interaction to lie somewhere between the feedback-loop limit and the equivalent circuit of section III. In order to perform calculations in the general case, we need to evaluate quantities such as  $\langle A^\dagger_{(n)}A_{(m)} \rangle$ for which the wave-packets might only partially overlap. We use the method of Rohde, Maurer and Silberhorn (RMS) \cite{ROH07} for decomposing a general wave-packet into components that are either perfectly matched or completely orthogonal to some mode of interest. As an example, let the mode of interest be $A_{(n)}$. According to RMS, we can always define a complete orthonormal set of functions $\{ A^{(i)}_{n} \}$ such that $A^{(0)}_{n} \equiv A_{(n)}$ and all the other modes with $i \neq 0$ are orthogonal to the selected mode. Then we can decompose any other mode, say $A_{(m)}$, as:
\begin{equation}\label{rms}
 \begin{aligned}
 A_{(m)} &= \lambda_0 A^{(0)}_{n} + \sum^{\infty}_{i \neq 0} \lambda_i A^{(i)}_{n}  \\
                            &\equiv \lambda_0 A^{(0)}_{n} + \sqrt{1-{\lambda_0}^2} \bar{A}_{n}     
 \end{aligned} 
\end{equation}
where the operator $\bar{A}_{n}$ contains the accumulated orthogonal modes, and
\begin{equation}
 \lambda_i \equiv \left[ A_{G,J^{(m)}},A^{\dagger (i)}_{n} \right] \, .
\end{equation}
Note that:
\begin{equation}
 \lambda_0 = \left[ \op{A}_{(m)}, \op{A}^\dagger_{(n)} \right]\, ,
\end{equation}
which is simply the commutator $C_{n,m}$ of (\ref{commJ}).

\subsection{Energy conservation}
Another general result that will prove useful is the expectation value $\bk{\op{A}'^\dagger \op{A}'}$, for any input state. Using (\ref{aoutbs2}) and separating into matched and orthogonal parts as outlined above, we find:
\eqn{\bk{\op{A}'^\dagger \op{A}'} &=& X\bk{\op{A}^\dagger \op{A}}+Y\bk{\op{A}^\dagger} \bk{\op{A}}, \nonumber \\
X &\equiv& \eta^2 \sum \limits^{\infty}_{m,n=0} e^{i(n-m)\phi}\sqrt{1-\eta}^{(n+m)}C_{n,m}+(1-\eta), \nonumber \\
Y &\equiv& \eta^2 \sum \limits^{\infty}_{m,n=0} e^{i(n-m)\phi}\sqrt{1-\eta}^{(n+m)} \sqrt{1-|C_{n,m}|^2} \nonumber \\
&-&\eta \sum \limits^{\infty}_{n=0}e^{i(n+1)\phi}\sqrt{1-\eta}^{(n+1)} C_{n,-1}-H.c.
}
The limits of the summations are not easy to determine analytically; however, because they converge exponentially, we can approximate them to arbitrary accuracy by truncating after an appropriate number of terms. After doing this, we find that $X\approx 1$ and $Y\approx 0$, leading to the result:
\eqn{\label{conserv} \bk{\op{A}'^\dagger \op{A}'}=\bk{\op{A}^\dagger \op{A}},}
regardless of the choice of input state $\op{A}$ and independent of the overlap $C_{n,m}$. This implies that the average number of particles is always conserved by the CTC evolution (\ref{aoutbs2}). Our model of the beamsplitter interaction therefore satisfies global energy conservation for all parameter choices, which is an important check of consistency.

\subsection{Numerical results for different input states}
For the coherent state, as remarked in section III, there is no entanglement produced by the beamsplitter and the evolution is described by standard quantum mechanics, by the application of a phase shift $e^{i \Phi(\eta,\phi)}$. We might then expect no changes as we smoothly go to the limit of a feedback loop, for which $\kappa \rightarrow 0$. Performing the calculations for different $\kappa$ using the general evolution (\ref{aoutbs2}), our expectations are confirmed: we obtain the same results as (\ref{coh2}), independently of the overlap $C_{n,m}$.

\begin{figure}[!htbp]
\includegraphics[width=8.6cm]{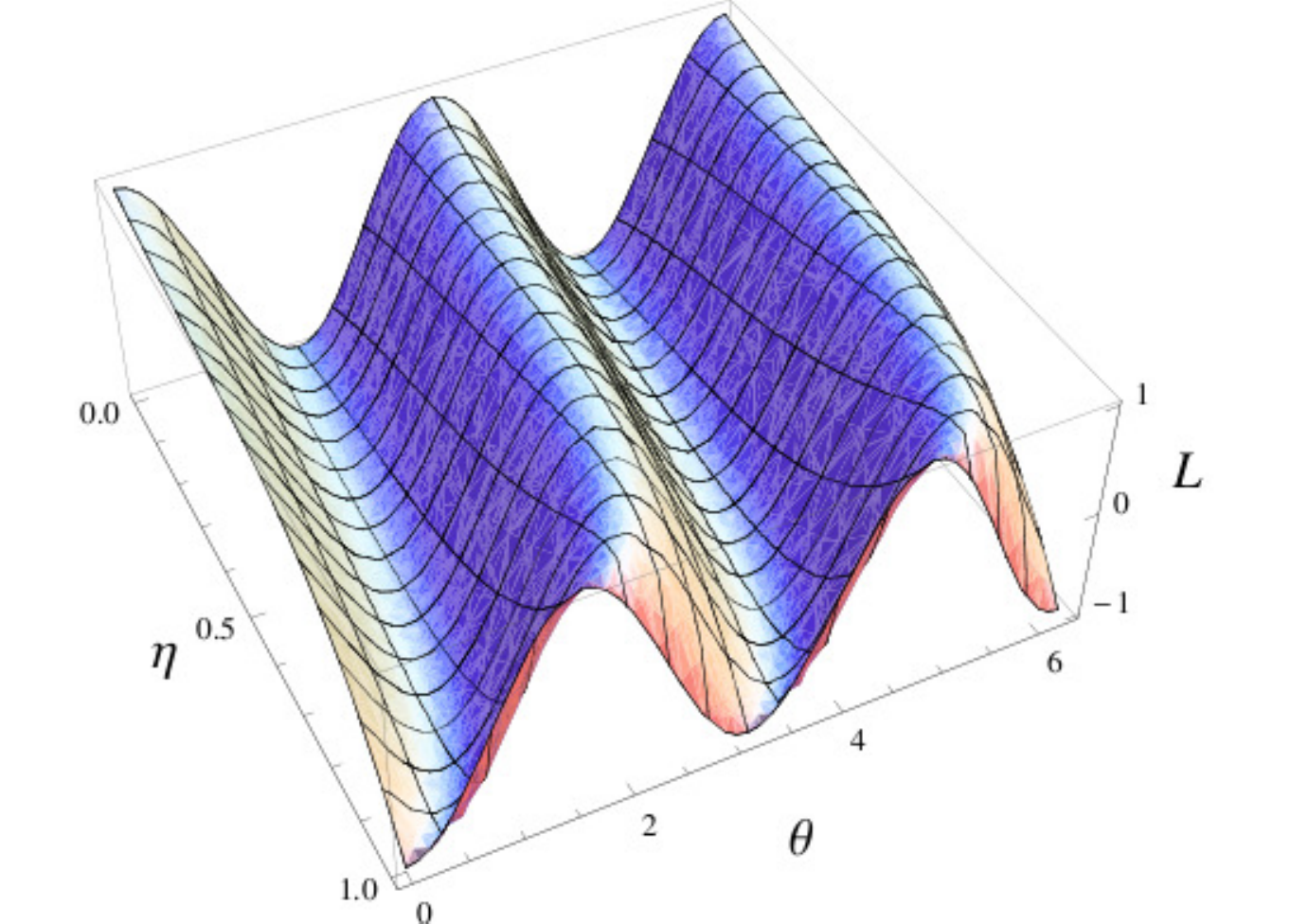}
\caption{(Colour online). Plot of $L(\frac{\pi}{2},\theta,\eta)$ for $\kappa=0$. We see that $|L(\frac{\pi}{2},\eta)|_{max}=1$ for any value of $\eta$, indicating that there is no loss of coherence due to the CTC. The overall effect is of a phase shift and a corresponding rotation of the squeezing angle. }
\label{kap0}
\end{figure}

\begin{figure}[!htbp]
\includegraphics[width=8.6cm]{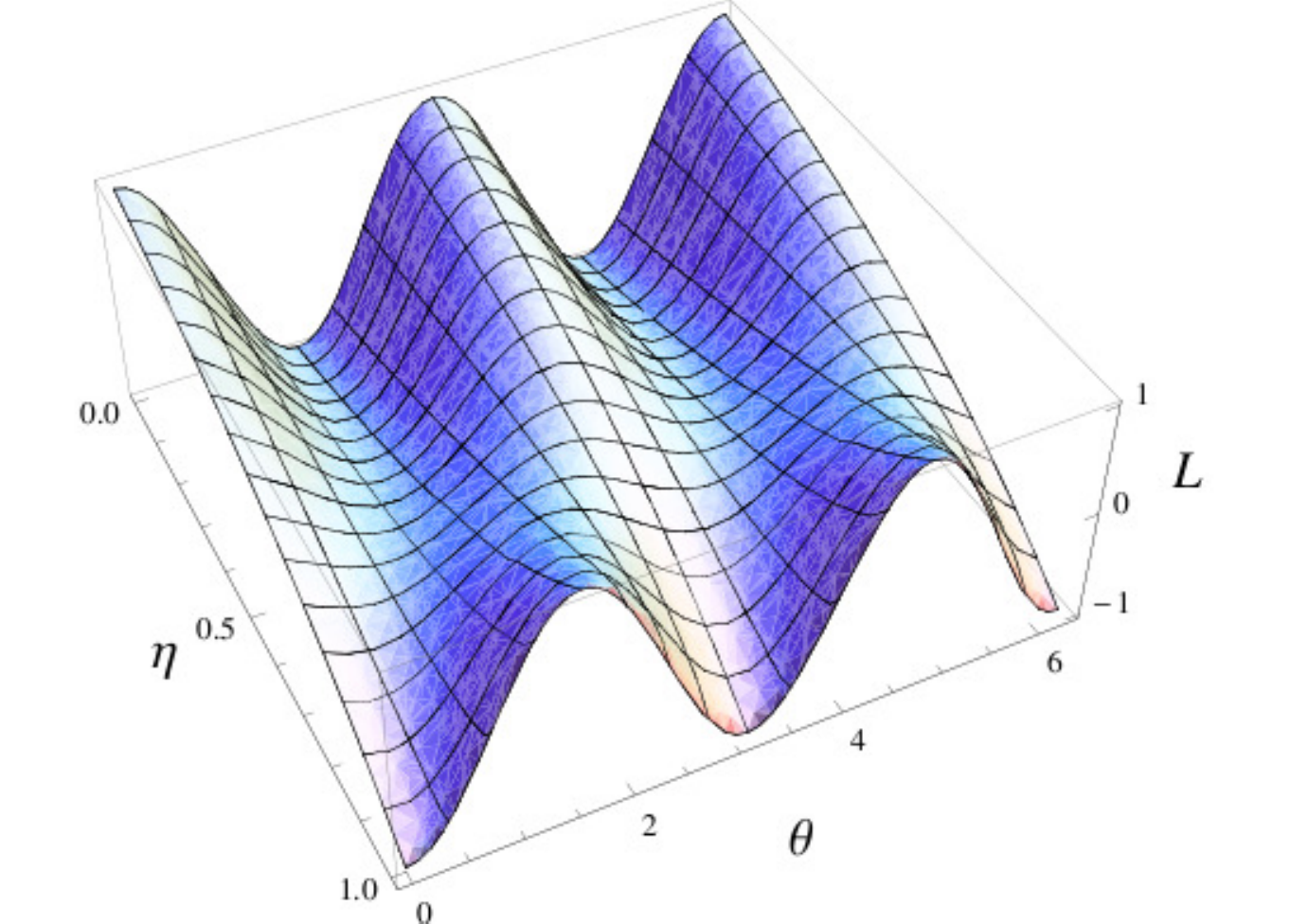}
\caption{(Colour online). Plot of $L(\frac{\pi}{2},\theta,\eta)$ for $\kappa=1$. Now $|L(\frac{\pi}{2},\eta)|_{max}<1$ for intermediate values of $\eta$, signifying some thermalisation due to the CTC interaction.}
\label{kap1}
\end{figure}

\begin{figure}[!htbp]
\includegraphics[width=8.6cm]{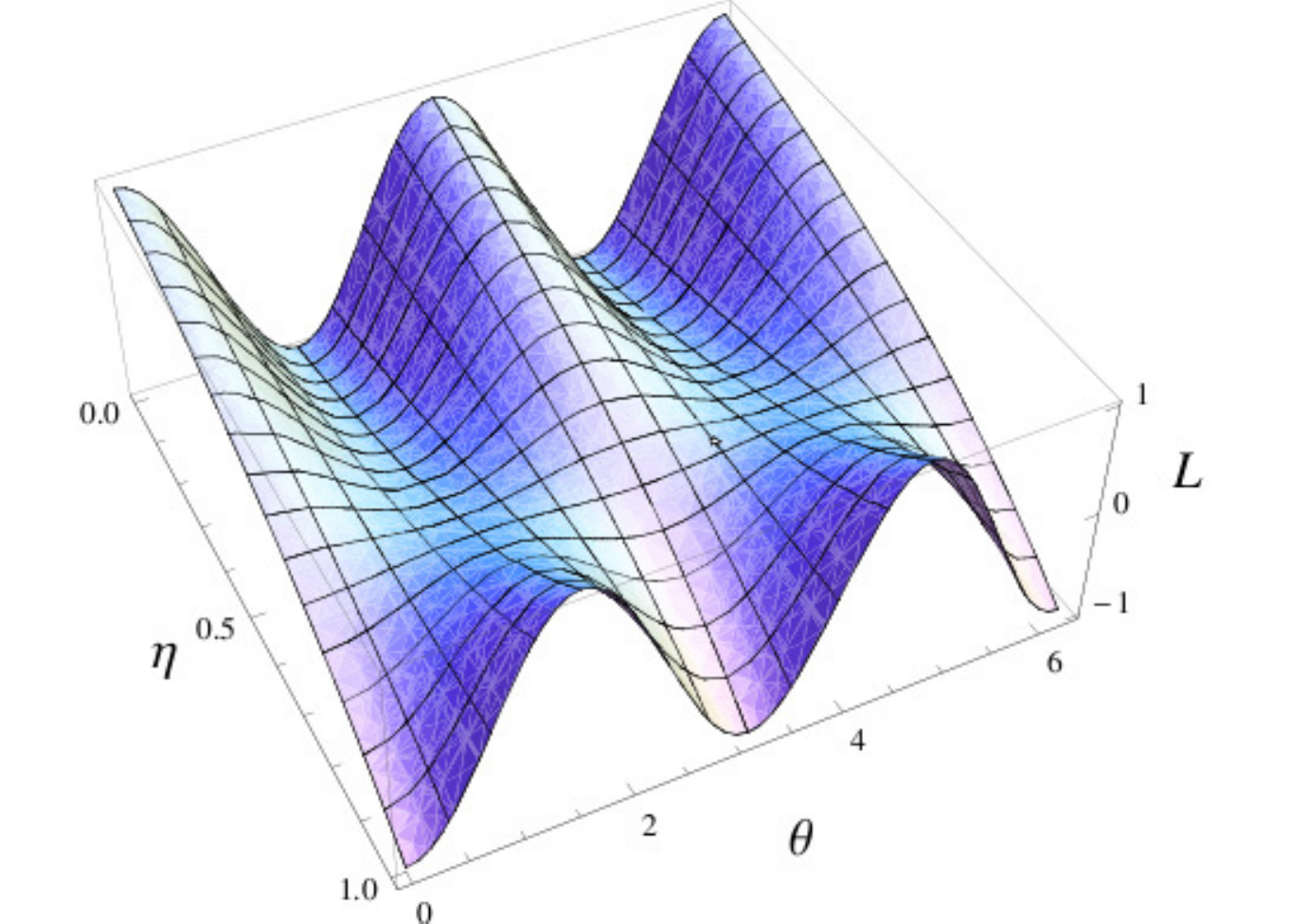}
\caption{(Colour online). Plot of $L(\frac{\pi}{2},\theta,\eta)$ for $\kappa=100$. The decoherence due to the CTC is at its maximum and we recover the equivalent circuit limit of section III (compare this graph to Fig.(\ref{phi2})). We find $|L(\frac{\pi}{2},\eta)|_{max}=0$ when $\eta=\frac{2}{3}$. }
\label{kap2}
\end{figure}

For the squeezed vacuum, only the quadrature variances differ from the results in (\ref{squeezed}). This is expected, since it is the variances that exhibit the effects of entanglement and decoherence due to the CTC. Performing the numerical calculations, we find that the function $L(\phi,\theta,\eta)$ now depends also on the parameter $\kappa$, such that $|L(\phi,\eta)|_{max} \rightarrow 1$ when $\kappa \rightarrow 0$, consistent with the feedback-loop limit (\ref{aoutstd}). In this limit, the phase shift results in a rotation of the squeezing angle as shown in Fig.\ref{kap0} as a function of $\eta,\phi$, but there is no noise as there is no coupling to the CTC. Plots of $L$ at $\phi=\pi/2$ (the value of $\phi$ for which the noise is maximised), for $\kappa=1$ and $\kappa=100$ are shown in Figs.\ref{kap1}-\ref{kap2} (colour online). As $\kappa$ increases, the noise (as measured by the amount of damping along $\theta$) increases until we obtain the equivalent circuit limit.

\begin{figure}[!htbp]
\includegraphics[width=8.6cm]{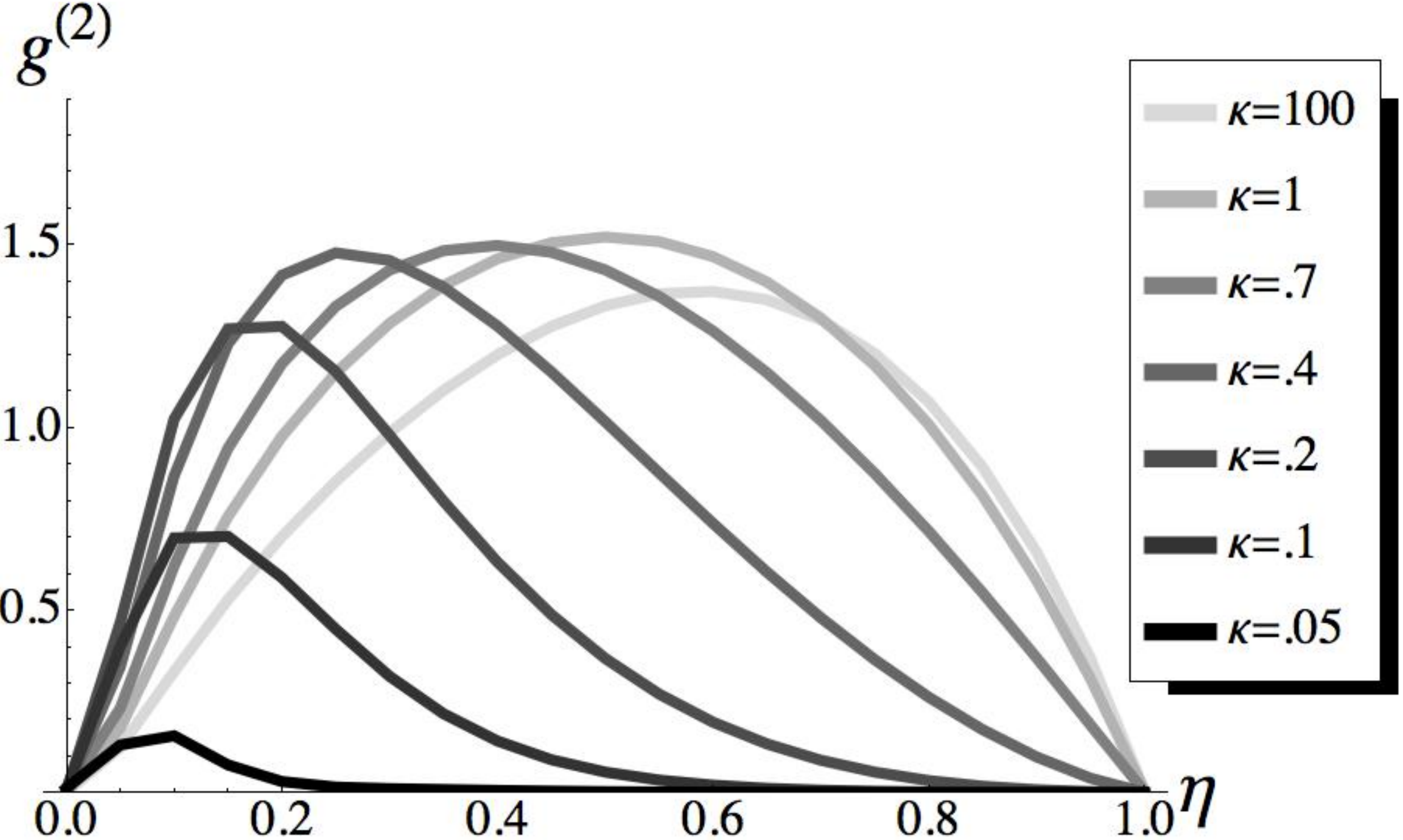}
\caption{$g^{(2)}$ as a function of $\eta$ for the output state when the input is a single photon. The graphs represent decreasing values of $\kappa$ (light to dark). For $\kappa \rightarrow \infty$ we obtain the result of the equivalent circuit, but as $\kappa \rightarrow 0$ the curve flattens out to zero, recovering the noiseless feedback-loop limit.}
\label{g2}
\end{figure}

Finally, we turn to the single photon state. It follows from the result (\ref{conserv}) that $\bk{\op{A}'^\dagger \op{A}'}=1$, i.e. the average photon number at the output is the same as for the input, in this case `1'. The interesting quantity is the probability of detecting photon numbers greater than 1, characterised by the second order correlation function $g^{(2)}$. To perform the calculation of $g^{(2)}$ for the equivalent circuit in section III, we made use of the identity (\ref{nmpq}), which was derived using the sharp commutator (\ref{comm2}). As we are now using the generalised commutator (\ref{commJ}), we need to derive a new identity for the term $\sum \limits_{n,m,p,q=0}^{\infty} \bk{\op{A}^\dagger_n \op{A}^\dagger_m \op{A}_p \op{A}_q}$ where $\op{A}$ is a single-photon mode described by the statistics (\ref{identities}). Using the RMS decomposition (\ref{rms}), we find:

\eqn{&\sum \limits_{n,m,p,q=0}^{\infty}& (\sqrt{1-\eta})^{n+m+p+q} \, \bk{\op{A}^\dagger_n \op{A}^\dagger_m \op{A}_p \op{A}_q} = \nonumber \\
&\sum \limits_{n,m,p,q=0}^{\infty}& (\sqrt{1-\eta})^{n+m+p+q} \, \bk{\op{A}^\dagger \op{A}}^2 \nonumber \\
 &\times & \left( C_{n,q} C_{m,p} + C_{n,p} C_{m,q} - 2 C_{n,m}C_{n,p}C_{n,q}  \right) \nonumber \\
}
(compare to (\ref{nmpq})). Using this result, and truncating the summations at an appropriate cutoff, we obtain graphs of $g^{(2)}$ as a function of $\eta$ for the different values of $\kappa$, shown in Fig.\ref{g2}.

The graphs shows that for $\kappa>0$, there is some probability of detecting photon numbers greater than 1 at the output. The input photon might disappear and nothing come out, or two or more photons might emerge, although the average photon count must remain 1 according to (\ref{cons}). The noise is nonzero only for intermediate values of $\eta$, corresponding to the region in which entanglement is created by the beamsplitter (again, there is no entanglement and hence no noise when we have perfect transmission or reflection). The shape of the distribution displays an asymmetry - this can be accounted for by noting that the physical circuit is itself asymmetric, because it takes two reflections for the photon to escape the CTC, but only one transmission. Indeed, if we were to alter the beamsplitter convention by swapping the outputs, we would obtain a mirror-reflection of $g^{(2)}$ around $\eta=0.5$. As $\kappa \rightarrow 0$, we approach the result $g^{(2)}=0$, which can also be obtained analytically from (\ref{aoutstd}). This limit corresponds to an effective decoupling from the CTC, so the noise vanishes and we obtain exactly one photon out with certainty. 

\subsection{The role of curvature}
Earlier, we remarked that our generalised model is formally identical to that found in \cite{RAL09}. There it was conjectured that such a generalised model would be compatible with CTC interactions - a claim that is confirmed by the analysis in this paper. In particular, in that work it was suggested that any space-time curvature, not just that due to a CTC, should be described by a model of the sort that we have introduced here. This would have implications for entangled particles in gravitational settings that could then be tested experimentally; this is made possible by the inherent non-linearity of the theory. It is tempting to disregard such a theory in favour of one which reduces to ordinary quantum mechanics in the absence of CTCs; however, the physical arguments we have made here seem to oppose that view. In particular, the equivalence that we have been led to postulate must hold between the spatial properties of the wave-packet and its interaction with the CTC implies that introducing curvature into our model would lead to it becoming inextricably linked with the extra degree of freedom; this is an interesting avenue for future research.

\section{Conclusions}
We have seen that it is possible to define a field theory that is consistent with Deutsch's model for quantum evolution on a CTC by applying field modes to the rails of the equivalent circuit; this model introduces an extra degree of freedom in order to achieve consistency with the Deutsch model. We used this model to perform calculations of the output for a coherent state, a squeezed state and a single-photon state interacting with a CTC on a beamsplitter.
Based on physical considerations, we postulated that the extra degree of freedom was related to the elapse of an affine parameter along the world-line of the time-traveling particle, which led us to a modified field theory capable of describing wave-packets comparable in size to the CTC itself.
Using this generalised model, we showed that it is possible to smoothly tune out the CTC by making the input modes much longer than the CTC. In this limit we found that the circuit reduces to a feedback loop with zero delay time, as described by normal quantum optics.

\begin{acknowledgements}
We thank A. Birrell for useful discussions. This work was supported by the Australian Research Council.
\end{acknowledgements}

\bibliographystyle{apsrev4-1}

\end{document}